\documentclass[11pt]{article}

\usepackage{pb-diagram}
\usepackage{latexsym}
\usepackage{amsfonts}
\usepackage[usenames,dvipsnames]{xcolor}
\usepackage{graphicx,amssymb,amsmath,epsfig}
\usepackage[english]{babel}
\usepackage{graphicx}
\usepackage{dcolumn}
\usepackage{bm}
\usepackage{setspace}
\usepackage{amsmath}
\usepackage{dynkin-diagrams}
\usepackage{tikz}
\usepackage{pgfplots}

\definecolor{myblue}{rgb}{0,0,0.8}
\usepackage[colorlinks=true
,urlcolor=myblue	
,anchorcolor=myblue
,citecolor=myblue
,filecolor=myblue
,linkcolor=myblue
,menucolor=myblue
,pagecolor=myblue
,linktocpage=true  
]{hyperref} 

\definecolor{green}{RGB}{0, 130, 0}
\definecolor{grey}{RGB}{90, 90, 90}

\catcode`\@=11
\def\marginnote#1{}

\newcount\hour
\newcount\minute
\newtoks\amorpm
\hour=\time\divide\hour by60 \minute=\time{\multiply\hour by60
\global\advance\minute by-\hour}\edef\standardtime{{\ifnum\hour<12
\global\amorpm={am}%
        \else\global\amorpm={pm}\advance\hour by-12 \fi
        \ifnum\hour=0 \hour=12 \fi
        \number\hour:\ifnum\minute<10
0\fi\number\minute\the\amorpm}}
\edef\militarytime{\number\hour:\ifnum\minute<10 0\fi\number\minute}

\def\draftlabel#1{{\@bsphack\if@filesw {\let\thepage\relax
   \xdef\@gtempa{\write\@auxout{\string
      \newlabel{#1}{{\@currentlabel}{\thepage}}}}}\@gtempa
   \if@nobreak \ifvmode\nobreak\fi\fi\fi\@esphack}
        \gdef\@eqnlabel{#1}}
\def\@eqnlabel{}
\def\@vacuum{}
\def\draftmarginnote#1{\marginpar{\raggedright\scriptsize\tt#1}}
\def\draft{\oddsidemargin -.5truein
        \def\@oddfoot{\sl preliminary draft \hfil
        \rm\thepage\hfil\sl\today\quad\militarytime}
        \let\@evenfoot\@oddfoot \overfullrule 3pt
        \let\label=\draftlabel
        \let\marginnote=\draftmarginnote

\def\@eqnnum{(\theequation)\rlap{\kern\marginparsep\tt\@eqnlabel}%
\global\let\@eqnlabel\@vacuum}  }

\def\numberbysection{\@addtoreset{equation}{section}
        \def\theequation{\thesection.\arabic{equation}}}

\def\underline#1{\relax\ifmmode\@@underline#1\else
 $\@@underline{\hbox{#1}}$\relax\fi}

\catcode`@=12 \relax

\numberbysection

\topmargin by -\headsep
\def\nonu{\nonumber}
\def\br{\begin{eqnarray}}
\def\er{\end{eqnarray}}
\def\lb{\lbrack}
\def\rb{\rbrack}

\def\[{\left[}
\def\]{\right]}

\def\lie{{\cal G}}
\def\a{\alpha}

\def\d{\delta}

\def\eps{\epsilon}

\def\g{\gamma}
\def\G{\Gamma}

\def\l{\lambda}

\def\o{\omega}
\def\O{\Omega}

\def\pa{\partial}

\def\s{\sigma}

\def\tp0{\Theta_{+}^{(0)}}
\def\tm0{\Theta_{-}^{(0)}}

\def\l{\lambda}

\def\nonu{\nonumber}

\def\bi{\begin{itemize}}
\def\ei{\end{itemize}}

\begin{document}


\title{Twisted  Affine Integrable Hierarchies and Soliton Solutions}
\begin{center}
{\Large\bf   Twisted  Affine Integrable Hierarchies and Soliton Solutions}
\end{center}
\normalsize
\vskip 1cm
\begin{center}
{Y.F. Adans}\footnote{\href{mailto:jYslar}{ysla.franca@unesp.br}},   J.F. Gomes\footnote{\href{mailto:francisco.gomes@unesp.br}{francisco.gomes@unesp.br}}, G.V. Lobo\footnote{\href{mailto:gabriel.lobo@unesp.br}{gabriel.lobo@unesp.br}}, and A.H. Zimerman\footnote{\href{mailto:a.zimerman@unesp.br}{a.zimerman@unesp.br}}\\[.7cm]

\par \vskip .1in \noindent
 {Instituto de F\'isica Te\'orica - IFT/UNESP,\\
Rua Dr. Bento Teobaldo Ferraz, 271, Bloco II,
CEP 01140-070,\\ S\~ao Paulo - SP, Brasil.}\\[0.3cm]

\vskip 2cm

\end{center}

\begin{abstract}
A systematic construction  of a class of integrable hierarchy is discussed in terms of the  twisted   affine   $A_{2r}^{(2)}$ Lie algebra.   The  zero curvature representation of the   time evolution equations are shown to be   classified according to its  algebraic structure and  according to its vacuum  solutions.
It is shown that  a class of models  admit  both zero and constant  (non zero)  vacuum solutions.  Another,  consists  essentially  of integral non-local equations  and can be classified into  two sub-classes, one admitting  zero vacuum and another  of constant, non zero vacuum solutions.  
    The two dimensional gauge potentials  in the vacuum plays a crucial ingredient and  are shown to be expanded in powers  of the  vacuum parameter $v_0$.
Soliton solutions are constructed  from vertex operators, which for the  non zero vacuum solutions, correspond to deformations characterized by  $v_0$.

  \end{abstract}

\vskip 1 cm
\section{Introduction}

The connection  between integrable hierarchies and the structure of affine Lie algebras  has provided a series of important achievements such as the systematic construction and classification of time evolution of nonlinear equations,   construction of soliton (multi) solutions, conservation laws, etc. (see for instance \cite{olive1},\cite{babelon} ).  
More recently  the theory of affine Lie algebras   has been   systematically employed  to derive and classify 
Bäcklund transformations (BT) \cite{ana1}-\cite{lobo2}.  These BT, in turn   were shown to naturally appear  in describing  integrable defects, in          the sense  that they   connect  two  field configurations of the same equation of motion without breaking the integrability.
Examples  in connection to sine-Gordon, Tzitzeica-Bollough-Dood  and  non relativistic models as well were considered, e.g., \cite{corrigan1}, \cite{corrigan2}, \cite{corrigan3}.  

In particular the  Tzitzeica-Bollough-Dood  (Tz-B-D) model corresponds to the relativistic  equation  based upon  the twisted $A_2^{(2)}$ affine algebra. The Tz-B-D for field $\phi$  can be obtained from the untwisted $A_2^{(1)}$  Toda model with fields $\phi_1, \phi_2$ identified together, i.e.,  $\phi \equiv  \phi_1=\phi_2$.
A sequence of nonlinear  time evolution equations  can then be derived from  the very same algebraic structure  of the Tz-B-D model  and zero curvature representation   to constitute   the $A_2^{(2)}$ twisted affine hierarchy.
In fact  such argument can be  generalized  mathematically to the affine  $A_{2r}^{(2)}$ case by identifying symmetries  of  Dynkin diagram  of $A_{2r}$  under  an automorphism  of  order two (see for instance \cite{olive-symm}).

It was shown in   \cite{corrigan2} and later in \cite{lobo1} that the  Bäcklund transformation  for the affine  $A_2^{(2)}$  hierarchy   has a  structure  much more elaborated than its untwisted  $A_2^{(1)}$  counterpart involving  an auxiliary external field.  An  elaborated study of  Bäcklund transformation  approach to  integrable defects for affine twisted  algebras  was  proposed in \cite{robertson}.
 Such transformation  was dubbed  type II Bäcklund transformation \cite{corrigan2} and is  a general feature of  twisted affine algebras  that underlines an entire  class of  nonlinear equations.

   In this paper we  extend the results of ref.   \cite{guilherme}  to  the case  of   affine twisted  algebras.   We first consider the $A_2^{(2)}$ affine algebra and construct two  sub-hierarchies associated to  some   {\it positive} and {\it negative} grade, constant  elements explained in section 3. 
  The former is shown  to admit   both, zero or  constant (non vanishing)  vacuum soliton solutions.    The negative sub-hierarchy, in turn  consists  essentially  of integral non-local equations  and can be classified into  two sub-classes, one admitting  zero vacuum and another  of constant, non zero vacuum solutions.

  An interesting feature that naturally  arises for  non zero vacuum solutions  is that the vacuum  structure for the two dimensional gauge potentials involves  both, the grading of the algebraic structure of the affine algebra  and  powers of a parameter  that characterizes  the vacuum.  
  The two  concepts   put together   defines a larger structure already encountered  within the {\it two loop Kac-Moody algebras }\cite{twoloop}, \cite{schwimmer}.

This paper is organized as follows.
In section 2 we  review the construction  of integrable hierarchies in terms of  decomposition of affine algebras and zero curvature representation.  Next, in section 3 we discuss  the construction of twisted  $A_2^{(2)}$ affine  algebra and  classify the possible sub-hierarchies in terms of graded algebraic elements and  in terms of their  vacuum  structure.  
In section 4 we discuss the dressing construction of   soliton solutions   and construct  the one soliton solutions for the $A_2^{(2)}$ hierarchy in terms of deformed  vertex operators.   Finally in section 5 we  discuss the general  construction for $A_{2r}^{(2)}$ twisted  affine algebra  and derive  a few simple examples. In section 6 we conclude and discuss further developments.


\section{Construction of Integrable Hierarchies}
Here we review the construction of integrable hierarchies in terms of a graded affine Lie algebra.
Consider an affine Lie algebra  $\hat \lie $ which can decomposed  according to a grading operator $Q$ as
\br 
\hat \lie  = \sum \lie_a, \qquad  [Q, \lie_a]\subset \lie_a, \qquad [\lie_a, \lie_b]\subset \lie_{a+b}, \quad a,b \in Z \label{grad}.
\er
Let $E\equiv E^{(1)}\in \lie_1$  be a semisimple  grade one element, which decomposes  the  $\hat \lie = {\cal K}  \oplus {\cal {M} }$, where  the kernel,  ${\cal K}$
 is defined to be
\br {\cal K} = \{  x\in {\cal {K}}, [x, E]=0 \}
\er
and $\cal {M}$ is its complement.
The Lax operator is defined as 
\br L = E + A_0
\er
where 
$A_0  \in {\cal M} \cap \lie_0$.

The integrable hierarchy is  constructed  from the zero curvature representation
\br
\lb \pa_x + E + A_0, \;\;  \pa_{t_{NM} }+ D^{(N)}+ \cdots + D^{(0)} + \cdots + D^{(-M)}\rb=0 \label{zcc}
\er
where $D^{(i)} \in \lie_i, \; $ and due to  the graded structure  (\ref{grad}) the  zero curvature  equations (\ref{zcc})  decomposes into
\br
\lb  E, D^{(N)} \rb &=&0 \label{h1} \\
\lb E, D^{(N-1)} \rb  +\lb A_0,D^{(N)}\rb + \pa_x D^{(N)} &=&0 \\
		& \vdots & \nonu \\
\lb E, D^{(-1)}\rb  +\lb A_0,D^{(0)}\rb  +\pa_x D^{(0)} -\pa_{t_{N,M}}A_0 &=&0 \label{h0} \\
		& \vdots & \nonu \\
\lb A_0, D^{(-M)}\rb  +\pa_x D^{(-M)} & =& 0 \label{h2}
\er
Notice that the $D^{(i)}, i>0$ can be solved recursively   starting from (\ref{h1}) downwards.  On the other hand the $D^{(-j)}, j>0$  are in turn solved  starting from (\ref{h2}) upwards until  we reach the zero grade  component  (\ref{h0}).  This  last  equation, (\ref{h0})     corresponds  to the time evolution equations  according to time $t_{NM}$ for fields  parameterizing $A_0 \in {\cal {M}} \cap \lie_0 $.
Two sub-hierarchies are of particular interest, namely,
\begin{itemize}
\item The{ \it  positive sub-hierarchy} is obtained  by setting  $D^{(-i)}=0, i>0$. Eqn(\ref{h1})  establishes  the possible values for $N$ since it  imposes the highest  grade element $D^{(N)}$   to belong to the kernel of $E$, $D^{(N)} \in {\cal K}_{E}$.

\item The {\it negative sub-hierarchy} obtained by taking $D^{(i)}=0, i\geq 0$.
An interesting particular example is the case where $N=0, \; M=1$, i.e., $t=t_{0,1}$ leading to 
\br
\lb A_0, D^{(-1)}\rb  +\pa_x D^{(-1)} & =& 0  \label{ls1}\\
\lb E, D^{(-1)}\rb    -\pa_{t_{0,1}}A_0 &=&0 \label{ls2}
\er
which can be solved  in  closed form for   general  Lie algebra  by a change of variables $ A_0 = B^{-1} \pa_x B$ and $D^{(-1)} = B^{-1} E^{(-1)}B, \quad  E^{(-1)}=E^{\dagger}$  for  some group element $B = e^ {\lie_0}$. The equation (\ref{ls1}) is automatically satisfied while (\ref{ls2}) leads to the Leznov-Saveliev eqn.
\br
\pa_t (B^{-1} \pa_x B) - [E, B^{-1} E^{(-1)} B]=0.  \label{ls3} 
\er
The above equation  correspond to the relativistic Toda  eqns. when  the  $(x, t_{0,1})$ variables are  identified    with  the light cone  coordinates, $ (z, \bar z) $.
\end{itemize}

Explicit examples were constructed for $A_2^{(1)}$  and its Bäcklund transformations  were  discussed in \cite{ana1}, \cite{ana2}. For $A_3^{(1)}$ and its generalization to   $A_n^{(1)}$  we  refer  to \cite{lobo1}, \cite{lobo2}.

\section{The Twisted $A_2^{(2)}$ Hierarchy}
In this paper we  shall discuss  the construction  of a class of integrable hierarchies connected to twisted  affine algebras    and  discuss its  classification in terms  of their possible  vacuum solutions.
Let us start with the simplest  case of  the $A_2$ Lie algebra with positive roots 
\br \a_1, \; \a_2 \quad {\rm and } \quad \a_1+\a_2
\er
and  $\s $,  an automorphism of order two such that $ \s(\a_1)=\a_2$, $\s^2 =1$.  Extending it to the Lie algebra,
\br \s (\a_1 \cdot H) = \a_2\cdot H, \qquad 
\s (E_{\a_1 }) =E_{ \a_2 }, \qquad 
\s (E_{\a_1+\a_2 }) =-E_{ \a_1+\a_2 }
\er
shows that  it  is consistent with $E_{\a_1+\a_2 } =[ E_{\a_1},E_{\a_2 }]$.  
The  twisted affine $A_2^{(2)}$  algebra is constructed  by assigning  integer  affine indices to the even  subalgebra under $\s$, i.e., $T_a^{(m)}, \; \s (T_a)=T_a$   and  semi-integer  to the odd part,  $T_a^{(m+1/2)}, \; \s (T_a)=-T_a, \; \; m\in Z$.
Consider  now the grading operator  $Q=6d+ (\mu_1+\mu_2)\cdot H$, where $\mu_i, \;\; i=1,2$ are the fundamental weights which   decomposes  the affine  $A_2^{(2)}$  as follows
\begin{equation}
\begin{split}
   {\lie_{6m}}&=\{h_1^{(m)}+h_2^{(m)}\}\\
     {\lie_{6m+1}}&=\{E_{\alpha_1}^{(m)}+E_{\alpha_2}^{(m)},E_{-(\alpha_1+\alpha_2)}^{(m+1/2)}\}\\
      {\lie_{6m+2}}&=\{E_{-\alpha_1}^{(m+1/2)}-E_{-\alpha_2}^{(m+1/2)}\}\\
       {\lie_{6m+3}}&=\{h_1^{(m+1/2)}-h_2^{(m+1/2)}\}\\
        {\lie_{6m+4}}&=\{E_{\alpha_1}^{(m+1/2)}-E_{\alpha_2}^{(m+1/2)}\}\\
        {\lie_{6m+5}}&=\{E_{-\alpha_1}^{(m+1)}+E_{-\alpha_2}^{(m+1)},E_{+(\alpha_1+\alpha_2)}^{(m+1/2)}\}.
\end{split}
\end{equation}

The Lax operator is  then constructed according to the  graded decomposition given above,
\br
L = E+ A_0, \qquad  E= E_{\a_1}^{(0)} + E_{\a_2}^{(0)} + E_{-\a_1-\a_2}^{(1/2)}, \qquad A_0 = v(x, t) (\a_1+\a_2)\cdot H^{(0)}.
\er

The  {\it positive sub-hierarchy}  requires the   time evolutions  to be  associated to the kernel elements ${\cal K}$  (see eqn. (\ref{h1}) ) which in such case  are given  by 
either, 
\br
E^{(6n -1)} &=& E_{-\a_1}^{(n)} + E_{-\a_1}^{(n)} + E_{\a_1+\a_2}^{(n-1/2)} \in \lie_{6n-1}\qquad  {\rm or} \label{6n-1}  \\ 
E^{(6n +1)} &=&E_{\a_1}^{(n)} + E_{\a_1}^{(n)} + E_{-\a_1-\a_2}^{(n+1/2)} \in \lie_{6n+1} \label{6n+1}.
\er
The structure of the kernel is such that the   possible  time evolution equations are derived from  elements associated  to grades, either $N=6n-1$ or $N=6n+1$.
The simplest  model  for the positive sub-hierarchy is then associated  to $M=0, N=5$, i.e., $t= t_{0,5}\equiv t_5$  in the language of (\ref{zcc}) (or  $N=6n-1 =5$ in (\ref{6n-1})).
Solving eqns. (\ref{h1})-(\ref{h2})  we find for  the evolution equation (\ref{h0})
\begin{equation}
\begin{split}
      A_x=E+v (h_1^{(0)}+h_2^{(0)}),\qquad 
      A_{t_5}=D^{(5)} + D^{(4)} + D^{(3)} + D^{(2)} + D^{(1)} + D^{(0)} 
\end{split}
\end{equation}
where
\br
D^{(5)} &=&E^{(5)}=E_{-\a_1}^{(1)}+E_{-\a_2}^{(1)}+E_{(\a_1+\a_2)}^{(\frac{1}{2})} \\
D^{(4)} &=& v(E_{\a_1}^{(\frac{1}{2})}-E_{\a_2}^{(\frac{1}{2})}) \\
D^{(3)} &=&\frac{1}{3}(v^2+v_x)(h_1^{(\frac{1}{2})}-h_2^{(\frac{1}{2})}) \\
D^{(2)} &=& -\frac{1}{3}\partial_x(v^2+v_x)(E_{-\a_1}^{(\frac{1}{2})}-E_{-\a_2}^{(\frac{1}{2})})\\
D^{(1)} &=& -\frac{1}{9}[(v^2+v_x)^2-2\partial_x^2(v^2+v_x)]E^{(1)}-\frac{1}{3}[v_x(v_x-2v^2)+\frac{1}{2}\partial_x^2(v^2+2v_x)]E_{-(\a_1+\a_2)}^{(\frac{1}{2})}  \\
D^{(0)} &=&-\frac{1}{9}(v^5-5vv_x^2-5v^2v_{2x}+5v_xv_{2x}+v_{4x})(h_1^{(0)}+h_2^{(0)})
\er
yielding the following equation of motion,
\begin{equation}
    9v_{t_5}=-v_{5x}-5v_{3x}v_x+5v^2v_{3x}-5v_{2x}^2+20vv_xv_{2x}-5v^4v_x+5v_x^3.  \label{t5}
\end{equation}
The second  simplest model is constructed by setting $M=0, N=7$, i.e., $t= t_{0,7}\equiv t_7$  in the language of (\ref{zcc})(or  $N=6n+1 =7$ in (\ref{6n+1})).
Solving eqns. (\ref{h1})-(\ref{h0})  we find   solution for $D^{(i)}, i=1, \dots, 7$  given in the appendix and  find  for   the evolution equation (\ref{h0}),
\br
    81v_{t_7} &=&28 v_{x} v^6-42 v_{3x} v^4-336 v_{x} v_{2x} v^3+28 v_{x}{}^4+168 v_{x}{}^2 v_{3x}+21 \left(2 v_{3x} v_{x}-12 v_{x}{}^3+2 v_{2x}{}^2+v_{5x}\right) v^2 \nonu \\
   & +& 42 v\left(4 v_{2x} v_{x}{}^2+3 v_{4x} v_{x}+5 v_{2x} v_{3x}\right) +21 v_{x} \left(11 v_{2x}{}^2-v_{5x}\right)-3 \left(14 v_{3x}{}^2+21 v_{2x} v_{4x}+v_{7x}\right) \nonu \\
    \label{t7}
\er
It is interesting to notice that  the two equations (\ref{t5}) and (\ref{t7})  admit   both,  $i)$ {\it zero  vacuum},  $v=0$ or  $ii)$ {\it constant vacuum}, $v= v_0 \neq 0 $  soliton solutions. 
This is a general fact that can be extended to   all models within the positive sub-hierarchy.
We assume that  for either,  zero  or constant vacuum configurations,  the  two dimensional gauge potentials $A_x^{vac}$ and $A_{t_{N}}^{vac}$ are  {\it constant}  affine Lie  algebra elements.
In general, the  zero curvature representation for  the vacuum configuration,
\br
\lb \pa_x+A_x^{vac}, \pa_{t_N}+ A_{t_{N}}^{vac}\rb = \lb E+v_0(h_1^{(0)}+h_2^{(0)}), D^{(N)}_{vac} +  D^{(N-1)}_{vac} +\cdots + D^{(0)}_{vac} \rb = 0
\er
 leads to the following explicit dependence of  $A_{t_N}^{vac} $ in terms of $v_0$ 
\br 
A_{t_{6n\pm 1}}^{vac} = E^{(6n\pm 1)} + v_0 d^{(6n\pm 1 -1)}_{vac} + v_0^2 d^{(6n\pm 1 -2)}_{vac} + \cdots + v_0^{6n\pm 1 } d^{(0)}_{vac}, \label{a-vac}
\er
where $D^{(k)}_{vac }= v_0^{k-N} d^{(k)}_{vac}\in \lie_k $, $\;\; N=6n\pm 1$ and $d^{k-N}_{vac} $ does not depend upon $v_0$ { { Notice that (\ref{a-vac})  does not contain terms of single gradation, but several  of different gradations}}

We now discuss in general terms  the possible  time evolutions  for {\it negative  sub-hierarchies} and its corresponding vacuum solutions.  Two cases  are to be considered:
\begin{itemize}
\item  {\bf {Negative Sub-Hierarchy with  zero vacuum solution}, $v_0=0$}. The zero curvature  representation (\ref{h2})  for $v=0$  implies that $\lb E, D^{(-M)}_{vac} \rb =0$ since $\pa_x D^{(-M)}_{vac} =0$ and hence $ D^{(-M)}_{vac} \in {\cal {K}}$.  
It then follows from (\ref{6n-1}) and  (\ref{6n+1})  that $N=0$ and either $M=6n-1$ or $M=6n+1$.

Following  the general construction  for the negative sub-hierarchy for $N=0, \; \; M=1, \quad (t=t_{-1})$  we find from (\ref{ls3}) the Tzitzeica (Bullough-Dodd) model,
\br 
\pa_{t}\pa_x \phi = e^{2\phi} - e^{-\phi} \label{tzi}. \label{tzit}
\er
where $B= e^{\phi (\a_1+\a_2)\cdot H}, \quad  \phi =\int ^x v(y)dy \equiv d^{-1} \phi$.
The Tzitzeica model  (\ref{tzit}) corresponds to the simplest model  ($N=0, \; M=1 $)  within the negative sub-hierarchy and it is clear that it  admits  only zero vacuum solution, i.e., $\phi_{vac}=0$.
Other models within this subclass  are associated to elements of the Kernel with  either $M=6n-1$ or $M=6n+1$.

\item { \bf {Negative Sub-Hierarchy with constant vacuum solution}},  $v_0\neq 0$.
The zero curvature  for the vacuum  configuration in this case reads,
\br
\lb E+v_0(h_1^{(0)}+h_2^{(0)}), D^{(-M)}_{vac} +  D^{(-M+1)}_{vac} +\cdots + D^{(-1)}_{vac} \rb = 0 \label{nega-vac}
\er
and implies that  $A_{t_{-M}}^{vac}$ has the following $v_0 \neq 0$ dependence, 
\br
A_{t_{-M}}^{vac}= d^{(-M)}_{vac} +  v_0^{-1}d^{(-M+1)}_{vac} + \cdots + v_0^{-M+2}d^{(-2)}_{vac} + v_0^{-M+1}d^{(-1)}_{vac}, \qquad  d^{(-1)} _{vac}\in {\cal {K}}. \label{34}
\er

Starting from the lowest  projection of (\ref{nega-vac}), namely,
\br
\left\lb v_0(h_1^{(0)}+h_2^{(0)}), D^{(-M)}_{vac} \right\rb =0 \label{vac1}
\er
implies that  for  $v_0\neq 0$, eqn. (\ref{vac1}) allows two possibilities, either 
\br
D^{(-M)}_{vac} =a \left(h_1^{(-m)} + h_2^{(-m)} \right) \quad {\rm {or}} \quad D^{(-M)}_{vac} =
b \left(h_1^{(-m+1/2)} - h_2^{(-m+1/2)} \right), \label{nega-v} 
\er 
and hence  $M=6m$ or $M=6m-3$ respectively.

The   simplest model within the negative  sub-hierarchy  sector is obtained  from (\ref{nega-v}) for $M=3$, i.e.,
for  $t= t_{-3}$.  Solving (\ref{h2})-(\ref{h0}) we find
\br
        D^{(-1)}&=&-3e^{d^{-1}v} d^{-1}\left(e^{-2d^{-1}v}d^{-1}(e^{d^{-1}v})\right)(E_{-\alpha_1}^{(0)}+E_{-\alpha_2}^{(0)})+6e^{-2d^{-1}v} d^{-1}\left(e^{d^{-1}v}d^{-1}(e^{d^{-1}v})\right)E_{\alpha_1+\alpha_2}^{(-\frac{1}{2})} \nonu \\
        D^{(-2)}&=&
3e^{-d^{-1}v}d^{-1}(e^{d^{-1}v})(E_{\alpha_1}^{(-\frac{1}{2})}-E_{\alpha_2}^{(-\frac{1}{2})}) \nonu \\
        D^{(-3)}&=&h_1^{(-\frac{1}{2})}-h_2^{(-\frac{1}{2})},   \nonu
         \er
where $ d^{-1} f = \int^xf(y)dy $.  It leads to 
 the following non-local equation of motion,
\begin{equation}
    - {{1}\over {3}}v_{t_{-3}}=e^{d^{-1}v} d^{-1}\left(e^{-2d^{-1}v}d^{-1}(e^{ d^{-1}v})\right)+2e^{-2d^{-1}v} d^{-1}\left(e^{d^{-1}v}d^{-1}(e^{ d^{-1}v})\right). \label{t-3}
\end{equation}
It is clear that eqn.  (\ref{t-3}) only admits  {\it non zero vacuum } solution $v=v_0\neq 0$ as can easily be checked if we denote $d^{-1}v_0 = v_0 x$,
\begin{equation}
    \begin{split}
0&=e^{v_0x} d^{-1}\left(e^{-2v_0x}d^{-1}(e^{ v_0x})\right)+2e^{-2v_0x} d^{-1}\left(e^{v_0x}d^{-1}(e^{ v_0x})\right)\\
&=e^{v_0x}d^{-1}\left(\frac{1}{v_0}e^{-v_0x}\right)+2e^{-2v_0x}d^{-1}\left(\frac{1}{v_0}e^{2v_0x}\right)\\
&=e^{v_0x}\left(-\frac{1}{v_0^2}e^{-v_0x}\right)+2e^{-2v_0x}\left(\frac{1}{2v_0^2}e^{2v_0x}\right) =0
    \end{split}
\end{equation}

It can be also checked that  $v_0 =0$   {\it does not}  satisfy  (\ref{t-3}) since $d^{-1} v_0|_{v_0=0} = c =$ constant.

Another example  of  constant vacuum solution  is to consider $M=6$ to obtain 
\begin{equation}
   v_{t_{-6}}=e^{d^{-1}v} d^{-1}\left(e^{-2d^{-1}v}d^{-1}W \right)
   +2e^{-2d^{-1}v} d^{-1}\left(e^{d^{-1}v}d^{-1}W\right)    \label{t-6}
   \end{equation}
where
\br
W=e^{d^{-1}v}d^{-1}\left( e^{d^{-1}v} d^{-1} ( e^{-2d^{-1}v}d^{-1} ( e^{d^{-1}v})+2 e^{d^{-1}v}d^{-1}(e^{-2d^{-1}v}) ) \right)
\er
It is straightforward    to verify that  $W=0$ for $v=v_0 \neq 0$,  $ d^{-1} v_0 = v_0 x$ and henceforth  is solution of (\ref{t-6}). Also, for $v_0=0$  it is clear that $W\neq 0$  and it is not solution for (\ref{t-6}).
 

\end{itemize}
\section{Dressing and the Construction of Soliton Solutions}

In this section we  discuss  the systematic construction  of soliton solutions  from the dressing formalism. 
The zero curvature  representation  implies the  two dimensional  gauge potentials  written in a  pure gauge form, in particular for the vacuum configuration with $ v_0 \neq 0$,   
\br
A^{vac}_x(v_0) =T_0^{-1}\pa_x T_0, \qquad A_t^{vac}(v_0) =T_0^{-1}\pa_t T_0
\er
or  $T_0 = e^{(A_x^{vac}) x +(A_t^{vac}) t }$ {\footnote{ Notice that  for the vacuum configuration $ [A^{vac}_x , A^{vac}_t ]=0$.}}. Here $T_0$  denotes  a key  group element  describing the  different vacua possibilities.
Once the  vacuum configuration   is known,  a nontrivial configuration, $T=T_0 \Theta$ 
 can be  obtained  by  gauge transformation,
\br
A_{\mu} = \Theta^{-1} A_{\mu}^{vac} \Theta  + \Theta^{-1} \pa_{\mu} \Theta \label{theta} 
\er
The dressing method  connects the vacuum to  a nontrivial configuration by gauge transformation (\ref{theta}) \cite{babelon}.  In fact there are two solutions for $\Theta$,
\br
 \Theta_+ = e^{q(0)} e^{q(1)} e^{q(2)}\cdots  \qquad \rm {and } \qquad   \Theta_+ = e^{p(-1)} e^{p(-2)} \cdots \label{twothetas}
\er
$q(i) \in \lie_{-i}, \quad p(i)\in \lie_i$, which can be determined by substituting  (\ref{twothetas}) into (\ref{theta}).   In particular  $e^{q(0)} = B^{-1} e^{-\nu \kappa}$. As a direct consequence  $\Theta_-\Theta_+^{-1} = T_0^{-1} g T_0$  where $g$ is a constant group element.  It   
 induces the  general formula, 
\br
< i| B e^{\nu \kappa}|j> = <i |T_0^{-1} g T_0|j> ,\;\;  i,j=0, \cdots , rank \;\; \lie \label{dress}
\er
where $|i >$ denotes the highest weight state in the sense that $p(i)|j>=\; \; <j |q(i) =\;\;  0, \;\; i=1,2, \dots $ and    $\kappa $ is  the central term of the affine Kac-Moody Lie algebra $\hat \lie $ and $g$ is a constant group element. {\footnote{ Notice that, in order to introduce  highest weight  states  $|i>$  we have to  
 extend the  affine loop algebra to the central extended Kac-Moody algebra. As a consequence we  have introduced an extra field $\nu$ associated to the  central term $\kappa$}}   
Equation (\ref{dress})  relates  physical fields  parameterizing $B$,
 obtained by integrating $A_0 = B^{-1} \pa_x B$, explicitly in terms  of   space and time coordinates $(x,t)$  from $T_0$.
 
 In order to obtain explicit  soliton solutions   we choose 
 \br 
 g = e^{F(\g )}, \qquad [A^{vac}_{\mu}, F(\g )] = \l_{\mu}(\g ) F(\g ). \label{vertex}
 \er
$F(\g )$ is called vertex operator which, in general,  has the property of being nilpotent, i.e. $F(\g )^{k+1}=0, $ for some $k\in Z$.  It therefore follows from (\ref{vertex}) that
\br
T_0^{-1} g T_0 = e^{\rho(x,t;\g )F(\g )} =1 + \rho(x,t;\g ) F(\g ) + \cdots + {{1}\over {k!}} \rho(x,t;\g )^k F(\g )^{k}, \quad \rho(x,t;\g ) = e^{-\l_1(\g ) x - \l_N(\g ) t_N}.
\er
Soliton solutions are directly related to a particular choice of $g$.  In particular,   for $l$-soliton solution   we consider 
\br
g = e^{F(\g _1)}e^{F(\g _2)}   e^{F(\g _l)}, \qquad T_0^{-1} g T_0  =  e^{\rho _1F(\g_1)}e^{\rho_2F(\g_2)}   e^{\rho_lF(\g_l)}
\er
where $\rho_j = \rho(x,t;\g_j)$.

We should point out the  different  vacuum configurations induces  different vertex operators. Consider for instance the constant vacuum  configuration, $A_{\mu}^{vac} (v_0), v_0 \neq 0$    regarded as a deformation upon   
 $A_{\mu}^{vac} (v_0=0)$.  Such  deformation  induces  deformations  to the vertex operators (parameterized  in terms of $v_0$) and in turn to the soliton solutions (see for instance \cite{guilherme}).  
We now consider explicit solution  for the equations  derived  in the previous section.


\subsection{ $A_2^{(2)}$ Solitons}
Consider  the following  auxiliary quantities {\footnote{ Notice that these quantities contain terms  of different $Q$ gradations.  In order to have  all terms with same  grade we may  introduce a second loop, $v_0=w$   and define a generalized  
  grading  operator $\tilde Q = Q + w {{\pa}\over {\pa w}} $ as proposed  within  the  Two loop Affine algebra, see \cite{twoloop}, \cite{schwimmer}}}
\br
\Omega_{(6m+1)} &  \equiv & E_{\a_1}^{(m)} +  E_{\a_2}^{(m)} + E_{-\a_1-\a_2}^{(m+1/2)} +v_0(h_1^{(m)}+h_2^{(m)}),  \label{O}\\
\G_{(6m+5)} & \equiv &E_{-\a_1}^{(m+1)} +  E_{-\a_2}^{(m+1)} +E_{\a_1+\a_2}^{(m+1/2)} 
+ v_0( E_{\a_1}^{(m+1/2)} -E_{\a_2}^{(m+1/2)}) \nonu \\
&+& {{v_0^2}\over {3}} (h_1^{(m+1/2)}-h_2^{(m+1/2)}) \label{G}
\er
with the property that 
\br
\lb \Omega_{(6m+1)}, \G_{(6l+5)} \rb = {{6l+1}\over {2}} \kappa \;\d_{m+l+1,0}
\er
The  vacuum   for the two dimensional gauge potentials   can be shown to be written entirely in terms of  quantities (\ref{O}) and (\ref{G}), i.e.,
\br 
A_{t_1}^{vac}&=& A_x^{vac} =\O_{(1)}\\
      A_{t_5}^{vac}&=&\G_{(5)}-\frac{v_0^4}{3^2}\O_{(1)}\\
      A_{t_7}^{vac}&=&\O_{(7)}-\frac{v_0^2}{3}\G_{(5)}+\frac{4v_0^6}{3^4}\O_{(1)}\\
       A_{t_{11}}^{vac}&=&\Gamma_{(11)}-\frac{v_0^4}{3^2}\Omega_{(7)}+\frac{5v_0^6}{3^4}\Gamma_{(5)}-\frac{2^3{v_0}^{10}}{3^6}\Omega_{(1)} \\
       A_{t_{13}^{vac}}&=&\O_{(13)}-\frac{v_0^2}{3}\Gamma_{(11)}+\frac{2^2 v_0^6}{3^4}\Omega_{(7)}-\frac{7 v_0^8}{3^5}\Gamma_{(5)}+\frac{35 v_0^{12}}{3^8}\Omega_{(1)}    \\
	 & \vdots & \nonumber
\er
For the negative sub-hierarchy  with  constant vacuum, $v_0 \neq 0$,
\br
A_{t_{-6m+3}}^{vac} &=& 3v_0^{-2} \G_{(-6m+5)}, \\
A_{t_{-6m}}^{vac} &=& v_0^{-1} \O_{(-6m+1)}, \quad m=1,2, \dots 
\er 
and  for those with zero vacuum, $v_0 =0$,
\br
A_{t_{-6m+1}}^{vac }&=& E^{(-6m+1)}\\
A_{t_{-6m+5}}^{vac }&=& E^{(-6m+5)}, \quad m=1,2, \dots 
\er
A vertex operator  satisfying  (\ref{vertex}) 
for $v_0 \neq 0$   can then be constructed.  Consider
\br
 F(\g)&=\sum_{j}3^{3j}\left[(v_0-\g)((v_0+\g))\right]{}^{-2j}\left[ (2v_0-\g)(2v_0+\g)\right]^{-j}\eps \left\{  (h_1^{(j)}+h_2^{(j)})+a_1\kappa \delta_{j,0}\right. \nonu \\
 &\left. a_2(E_{\a_1}^{(j)}+E_{\a_2}^{(j)})+a_3E_{-(\a_1+\a_2)}^{(j+1/2)}+a_4(E_{-\a_1}^{(j+1/2)}-E_{-\a_2}^{(j+1/2)})+a_5(h_1^{(j+1/2)}-h_2^{(j+1/2)})\right.
 \nonu \\
 & \left. +a_6(E_{\a_1}^{(j+1/2)}-E_{\a_2}^{(j+1/2)})+a_7(E_{-\a_1}^{(j+1)}+E_{-\a_2}^{(j+1)})+a_8 E_{(\a_1+\a_2)}^{(j+1/2)}  \right\} \label{ver-v0}
\er

where $\eps $ is a free parameter and 
\br
        &a_1= -(3 \g)^{-1}(\g+v_0), \quad \quad a_2=\left[ \left(v_0-\g\right)\right]^{-1} ,\quad \quad a_3=2 (\g+2 v_0)^{-1},\\
       & a_4= -3  \g\left[\left(v_0-\g\right) \left(\g+v_0\right) \left(\g+2 v_0\right)\right]^{-1},\quad \quad
        a_5= -3 \left[\left(v_0-\g\right) \left(\g+v_0\right) \left(\g+2 v_0\right)\right]^{-1}\\&a_6=- 9 \left[\left(v_0-\g\right){}^2 \left(\g+v_0\right) \left(\g+2 v_0\right)\right]^{-1},\quad \quad
        a_7= -9[ \left(\g+2 v_0\right) \left(v_0^2-\g^2\right){}^2]^{-1},\\&a_8=18\left[\left(v_0-\g\right){}^2 \left(\g-2 v_0\right) \left(\g+v_0\right) \left(\g+2 v_0\right)\right]^{-1}.
  \er
It can be verified by direct calculation that  
\br
    \lb \O_{6m+1}, F(\g)\rb &=&(-3)^{-3m}\g  \left( \g^2-4 v_0^2\right){}^m \left(v_0^2-\g^2\right){}^{2 m} F(\g), \label{OO} \\
 \lb \G_{6m+5}, F(\g)\rb &=&(-1)^{-m} (3)^{-3 m-2} \g\left(\g^2-4 v_0^2\right){}^{m+1} \left(v_0^2-\g^2\right){}^{2 m+1}F(\g), \label{GG}
\er
and henceforth,
\br
    \lb A_{x}^{vac}, F(\gamma) \rb &=& \lambda_{1} F(\gamma)= \g F(\gamma) , \\
    \lb A_{t_j}^{vac}, F(\gamma) \rb &=&\lambda_{j} F(\gamma),
\er
where
 \br
   \lambda_{5}&=&\frac{1}{9}\g\left(\g^2-4 v_0^2\right) \left(v_0^2-\g^2\right) -\frac{v_0^4}{9}\g =\frac{1}{9} \left(5 v_0^2 \gamma^3-5 v_0^4 \gamma-\gamma^5\right),
   \label{l5}\\
   \l_7&=& \frac{1}{81} \left(-3 \g^7+21 \g^5 v_0^2-42 \g^3 v_0^4+28 \g v_0^6\right).\label{l7}  
\er
It therefore follows   for $j=5,7 $  that
\br
    \rho(t_5,x)&=&e^{-\lambda_1 x-\lambda_5 t_5}=e^{-\gamma x-\frac{1}{9} \left(5 v_0^2 \gamma^3-5 v_0^4 \gamma-\gamma^5\right) t_5}, \label{5} \\
    \rho(t_7,x)&=&e^{-\lambda_1 x-\lambda_7 t_7}=e^{-\gamma x-  \frac{1}{81} \left(-3 \g^7+21 \g^5 v_0^2-42 \g^3 v_0^4+28 \g v_0^6\right)t_7},  \label{7}
\er
 and similarly for higher  values of $j$.
For  the negative sub-hierarchy  with non zero vacuum, we find
 \br
  \l_{-3} &=&=\frac{3^2 \gamma}{v_0^2\left(\gamma-v_0\right)\left(\gamma+v_0\right)}, \label{l-3}  \\
 \l_{-6} &=&-\frac{3^3 \gamma}{v_0 \left(\gamma-2 v_0\right)\left(\gamma+2 v_0\right) \left[\left(\gamma- v_0\right)\left(\gamma+ v_0\right)\right]^2},
\label{l-6} 
\er
and 
\br
    \rho(t_{-3},x)&=&e^{-\gamma x-\frac{3^2 \gamma }{v_0^2\left(\gamma-v_0\right)\left(\gamma+v_0\right)}t_{-3}}, \label{-3} \\
    \rho(t_{-6},x)&=&=e^{-\gamma x+\frac{3^3 \gamma }{v_0 \left(\gamma-2 v_0\right)\left(\gamma+2 v_0\right) \left[\left(\gamma- v_0\right)\left(\gamma+ v_0\right)\right]^2}t_{-6}}, \label{-6}
\er
For $v_0 =0$, the vertex  operator (\ref{ver-v0}) becomes 
\br
F(\g)&=&\sum_{j}3^{3j}\gamma^{-6j} \epsilon \{(h_1^{(j)}+h_2^{(j)})-\frac{1}{3}\kappa \delta_{j,0}-\g^{-1}(E_{\a_1}^{(j)}+E_{\a_2}^{(j)})+2\g^{-1}E_{-(\a_1+\a_2)}^{(j+1/2)}\nonu \\
&+&3\g^{-2}(E_{-\a_1}^{(j+1/2)}-E_{-\a_2}^{(j+1/2)})
+3\g^{-3}(h_1^{(j+1/2)}-h_2^{(j+1/2)})\nonu \\
  &-&9\g^{-4}(E_{\a_1}^{(j+1/2)}-E_{\a_2}^{(j+1/2)})
  -9\g^{-5}(E_{-\a_1}^{(j+1)}+E_{-\a_2}^{(j+1)})\nonu \\
  &+&18\g^{-5} E_{(\a_1+\a_2)}^{(j+1/2)} \}. \label{v0}
\er

 It  coincides  with the vertex  constructed  in \cite{assis} for the Tzitzeica model where  $\g = \sqrt{3}z$
and (\ref{v0}) satisfy
\br
\lb  A_{t_{-6m+1}}^{vac }, F(\gamma)\rb& =&(-1)^{-m}{(3)}^{3m}\gamma^{-6m+1}F(\gamma), \label{3.38} \\
\lb A_{t_{-6m+5}}^{vac }, F(\gamma)\rb &=&(-1)^{-m+1}{(3)}^{3m-2}\gamma^{-6m+5}F(\gamma), \label{3.39} 
\er
leading to
\br
    \rho(t_{-6m+1},x)&=& e^{-\gamma x-(-1)^{-m}{(3)}^{3m}\gamma^{-6m+1} t_{-6m+1}}, \label{-6m+1} \\
    \rho(t_{-6m+5},x)&=& e^{-\gamma x+(-1)^{-m+1}{(3)}^{3m-2}\gamma^{-6m+5} t_{-6m+5}}. \label{-6m+5}
\er
Eqns. (\ref{5})-(\ref{7}),   (\ref{-3})-(\ref{-6})  and    (\ref{-6m+1})-(\ref{-6m+5}) establishes the  space-time dependence for each particular model.



\subsection{One Soliton Solution}

Here we explicitly  derive the one soliton solution for the  $A_2^{(2)}$ hierarchy.
  Propose the following ansatz,
\br 
 v(x, t_{-1}) = v_0 + \pa_x  \ln ( \tau_0/\tau_1 )
\er
where
\br
\tau_i =1+  \langle i |F |i\rangle \rho +  \langle i |F F  |i\rangle \rho^2,  \label{sol-tzit}
\er 
 $|i\rangle, \; i=0,1$ labels the  two fundamental weights of $A_2^{(2)}$  and $\rho(x,t_{-1}) $ is constructed from  eqns. (\ref{3.38}) and (\ref{3.39}) for $m=0$ and $m=1$ respectively,
\br
\rho (x,t_{-1}) = e^{-\g x -3 \g^{-1} t_{-1}}. \label{rho}
\er
The matrix elements  can be evaluated  in terms of representations of the affine $A_2^{(2)}$ Kac-Moody algebra  where (see for instance   \cite{assis}).   
\br
&(h_1^{(0)}+h_2^{(0)})| 0 \rangle = 0, \qquad & \kappa |0\rangle =2|0\rangle\nonu \\
&(h_1^{(0)}+h_2^{(0)})| 1 \rangle = 1|1\rangle, \qquad & \kappa |1\rangle = 2|1\rangle
\er
It therefore follows using vertex (\ref{ver-v0}) we find
\br
 c_0^{(1)}= \langle 0 |F | 0 \rangle &=& -{{2(\g +v_0) }\over {3\g}} \eps, \qquad c_0^{(2)}= \langle 0 |F^2 | 0 \rangle= {{(\g^2-4v_0^2)(\g +v_0)}\over {36 \g^2(\g-v_0)}} \eps^2 \nonu \\
c_1^{(1)}= \langle1  |F | 1\rangle &= & {{\g -2v_0}\over {3\g}}\eps , \qquad c_1^{(2)}=\langle 1 |F^2 | 1 \rangle= {{(\g-2v_0)^2}\over {36 \g^2}}\eps^2 \label{matrix}
\er
Notice that the matrix elements  (\ref{matrix}) are independent of space-time  and  hence  are  the same for all models within the hierarchy.
For other models of the $A_2^{(2)}$ hierarchy  with zero vacuum  ($v_0 =0$) we find 
the same functional  expression  for the tau-functions with $\rho (x, t_N) = e^{\g x + \l_{N}(\g )t_{N}}, \quad N=-1,3,\dots $ in the limit $v_0 \rightarrow 0$.
The matrix elements (\ref{matrix})  evaluated with either vertices  (\ref{ver-v0}) or (\ref{v0}) yields solution for all models within the hierarchy, i.e.,
\br
v(x, t_{k}) = \pa_x \phi (x, t_{k}) = v_0 + \pa_x  \ln \left( {{1+ c_0^{(1)} \rho(x,t_k)+ c_0^{(2)} \rho^2(x,t_k)} \over {1+ c_1^{(1)}\rho(x,t_k)+ c_1^{(2)}\rho ^2(x,t_k)}}\right) \label{solution}
\er
In order to check  consistency  of solutions (\ref{solution}) with eqns. of motion  let us consider for instance   $t_5$ eqns. (\ref{t5}) and propose
\br
&\tau _0(x,t)= 1+ \omega _1 \a \rho _1(x,t)+\gamma _1\a ^2  \rho _1(x,t){}^2,\\
&\tau _1(x,t)= 1+ \omega _2 \a \rho _1(x,t)+\gamma _2\a ^2  \rho _1(x,t){}^2
\er
with 
\br
\rho_1(t,x)=e^{f_1 x+g_1 t},
\er
Substituting in  (\ref{t5}), after  taking the least common  multiple  we find by  collecting powers of $\rho$. The lowest   power of $\rho $  leads to
\begin{equation}
    -5 f_1^3 v_0^2+5 f_1 v_0^4+f_1^5+9 g_1=0. \label{gamma1}
\end{equation}
If we set $f_1 =-\g $ we  find agreement with (\ref{rho}), i.e. $g_1 = -{{1}\over {9}} (5{v_0}^2 \g^3 -5{v_0}^4\g -\g^5)$.
Inserting in the coefficient of $\rho^2$ we find,
\br 
6 \gamma  (\gamma _2-\gamma_1) \left(\gamma -v_0\right)=\left(\omega _2-\omega _1\right) \left(\omega _1 \left(2 \gamma ^2-5 \gamma  v_0+2 v_0^2\right)+\omega _2 \left(4 \gamma ^2-\gamma  v_0-2 v_0^2\right)\right).
\er
Solving for $\g_1$ and inserting in the coeff. of $\rho^3$,
\br
 -36 \gamma _2 \gamma ^2 \left(\gamma -v_0\right)=\left(\omega _1-\omega _2\right) \left(\gamma -2 v_0\right) \left(\omega _2 \left(9 \gamma ^2+v_0 \left(\gamma -2 v_0\right)\right)+\omega _1 \left(3 \gamma -v_0\right) \left(\gamma -2 v_0\right)\right),
\er  
which can be solved for $\g_2$.
The coefficient of $\rho^4$ leads to 
\br
20\gamma ^3 \left(\omega _1-\omega _2\right){}^3 \left( \gamma -2 v_0\right){}^2 \left(17 \gamma ^2-5 v_0^2\right) \left(\omega _1 \left(\gamma -2 v_0\right)+2 \omega _2 \left(\gamma +v_0\right) \right) =0,
\er
and henceforth  gives the following non-trivial, ($w_1 \neq w_2$) solution,
\begin{equation}
    \omega _2= -\frac{\omega _1 \left(\g -2 v_0\right)}{2 \left(\g +v_0\right)}, \quad \gamma _1= \frac{\omega _1^2 \left(\g^2-4 v_0^2\right)}{16 \left(\g^2-v_0^2\right)}, \quad \gamma _2= \frac{\omega _1^2 \left(\g -2 v_0\right){}^2}{16 \left(\g+v_0\right){}^2} \label{hsol}
\end{equation}
The coefficients of higher powers of $\rho$  all vanish when  (\ref{hsol}) is taken into account.
In order to fit  with  solution (\ref{solution}), we set $f_1 =-\g $, $\a=\frac{\eps}{3\g}$ and $\o_1=-2 \left(\g+v_0\right)$ leading to
\br
g_1 = \frac{1}{9} \left(-5 v_0^2 \g^3+5 v_0^4 \g+\g^5\right),  \qquad \o_1 =c_0^{(1)}  \qquad \o_2 = c_1^{(1)} 
\er
and
\br
 \g_1 =c_0^{(2)} \qquad \g_2 = c_1^{(2)}.
\er




\section{Twisted Affine $A_{2r}^{(2)}$ Algebra}
We  shall now discuss  the construction of a class of integrable hierarchies based upon the twisted affine Lie algebra  $A_{2r}^{(2)}$.
In general, consider  the following  second order automorphism  for $A_{2r}$,
\br 
\s(\a_1)=\a_{2r}, \quad \s(\a_2)=\a_{2r-1}, \quad \cdots  \quad  \s(\a_{r-1} )=\a_{r+2}, \quad  \s(\a_r)=\a_{r+1}
\er
diagramatically illustrated  according to the Dynking diagram,
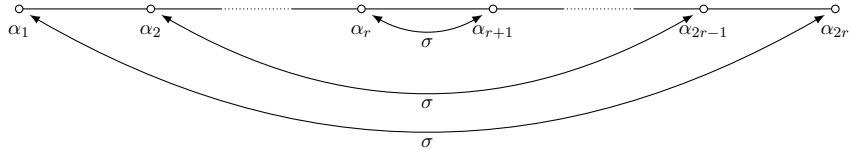
\begin{center}
\dynkin[edge length=1.75cm,
labels={\a_1,\a_2,\a_r,\a_{r+1},\a_{2r-1},\a_{2r}},
involutions={
1<below>[\s]6;
2<below>[\s]5;
3<below>[\s]4}
]{A}{oo.oo.oo}
\end{center}
Extending  to the Lie algebra  
we find,
\br 
\s (\a_i \cdot H) &=& \a_{2r-i+1}\cdot H, \quad i=1, \dots, r \nonu \\
\s (E_{\a }) &=&\zeta_{\a} E_{ \s(\a)}, \quad  \zeta_{\a} = \pm 1.
\er
E.g.,
\br \zeta _{\a_i}& =& +1,\quad  i=1, \dots, r, \nonu \\
\zeta _{\a_i+\a_{i+1}}& =& -1, \quad i=1, \dots, r-1, \nonu \\
	& \vdots  &      \nonu \\
\zeta _{\a_i+\cdots + \a_{j}}& =& (-1)^{j-i} ,\quad  i=1, \dots, r, \quad j=i, i+1, \dots, r.
\er
The   twisted affine   $A_{2r}^{(2)}$ algebra is constructed by assigning integer  affine indices to the even sub-algebra under $\s$, i.e.,
$T_a^{(m)}, \quad \s(T_a) = T_a$   and semi-integer indices to the odd part, $T_a^{(m+1/2)}, \quad \s(T_a) = -T_a$ , $m\in Z, a= 1, \dots,  \dim{A_{2r}}$ .

We now define  the grading operator (principal gradation) 
\br Q = 2 (2r+1)d + \sum_{i=1}^{2r} \mu_i \cdot H
\er
where $d$ is the derivation operator{ \footnote{ $d$ accounts for the  affine index , i.e., $[d, T_a^{(n)}]=nT_a^{(n)}$}} and $\mu_i, i=1, \dots, 2r$ are the fundamental weights 
of  the $A_{2r}$ Lie algebra. It therefore  follows that $Q$  induces a decomposition  of  $A_{2r}^{(2)} $
 into graded subspaces,
\br 
\lie_{2(2r+1)m+[0]}&=& \{(1+\s) h_i^{(m)}, \quad i=1, \dots, r \}   \label{g1}\\
\lie_{2(2r+1)m+[1]}&=& \{(1+\s)  E_{\a_i}^{(m)}, i=1,\dots, r, (1-\s) E^{(m+1/2)}_{-(\a_1+ \cdots + \a_{2r})} \} \\
\lie_{2(2r+1)m+[2]}&=& \{(1+\s)  E_{\a_i+\a_{i+1}}^{(m)}, \quad i=1,\dots, r-1, \nonu \\
& & (1-\s) E^{(m+1/2)}_{-(\a_1+\cdots + \a_{2r-1})}, \quad (1-\s) E^{(m+1/2)}_{-(\a_2+\cdots+ \a_{2r})}\} \\
	& \vdots & \nonumber \\
\lie_{2(2r+1)m+[{{1\over {2}}}2(2r+1)]}&=& \{(1-\s)  h_i^{(m+1/2)} , \quad i=1, \dots, r\} \\
\lie_{2(2r+1)m+ [{{1\over {2}}}2(2r+1)+1]}&=& \{(1-\s)  E_{\a_i}^{(m+1/2)}, i=1,\dots, r, (1+\s) E^{(m+1)}_{-(\a_1+\cdots+ \a_{2r})} \} \\
	& \vdots & \nonumber \\
\lie_{2(2r+1)m+[2(2r+1)-1]} &=& \{  (1+\s)  E_{-\a_i}^{(m)}, i=1,\dots, r,  (1-\s) E^{(m+1/2)}_{\a_1+\cdots+ \a_{2r}} \} \label{gn}
\er

The Lax operator is constructed by specifying a constant grade one operator
\br
E \equiv E^{(1)} = \sum_{i=1}^{2r}E_{\a_i}^{ (0)} + E_{-(\a_1 + \cdots + \a_{2r})}^{ (1/2)} \in \lie_{1}
\er
which, in turn  decomposes  the affine algebra $ \hat {\lie } = {\cal K} \oplus {\cal M}$ where ${\cal K} $ denotes the Kernel of $E$ such that  $ x\in {\cal K}, [E, x]=0$ and ${\cal M}$ is its complement.
The Lax operator is then  defined to be
\br 
L \equiv  E + A_0 = E + \sum_{i=1}^{r} v_i (x,t)(1+\s)h_i^{(0)} . 
\er


Following the algebraic data developed in    appendix B  we now illustrate  with   the       first few models for the $A_4^{(2)}$ hierarchy.
Let us redefine $v_1 \equiv v$ and $v_2 \equiv w$  for simplicity.
The simplest case is given by  $N=3$, yielding the  system of  equations of motion,
\br
        5{v}_{t_3}&=& -3v{}^2 \left({v}_{x}-2 w_{x}\right) +3v \left(4 wv_{x}-2w w_{x}-v_{2x}+w_{2x}\right)  \nonu \\
        &-&3 w \left(v_{2x}-w_{2x}\right)-3 v_{x}{}^2+3 w_{x}{}^2-3 w{}^2 v_{x}+2 v_{3x}-3 w_{3x} \label{v} \er
\br
5w_{t_3}&=&w^2 \left(6 w_x-9 v_{x}\right) +3v \left(6 wv_{x}-6ww_{x}-2 v_{2x}+w_{2x}\right) \nonu \\
       &+ &w\left(3 w_{2x}-6 v_{2x}\right)-6 v_{x}{}^2+3 w_x{}^2-3 v_{x} w_x+9 w_x v^2+3 v_{3x}-7 w_{3x} \label{w}
  \er
Clearly,  constant vacuum  $(v,w)=(v_0,w_0) \neq 0$,   as well as  zero vacuum   $(v,w)=(0,0) $, are  both solutions of (\ref{v})-(\ref{w}).

For the negative sub-hierarchy  the simplest  model  corresponds to $M=-1, \quad t = t_{-1}$,
\begin{equation}
    A_x=E^{(1)}+B^{-1}\partial_x B, \qquad  A_t=B^{-1}E^{(-1)}B,\qquad B= e^{(h_1+h_4)\,\phi_1(x,t)+(h_2+h_3)\,\phi_2(x,t)},
\end{equation}
where $E^{(-1)} = E^{\dagger}\in {\cal {K}}$  and $v = \pa_x \phi_1,\;\; w=\pa_x\phi_2$ leading   to  the equations of motion,
 \begin{gather}
    \partial_{xt}\phi_1=e^{2\phi_1-\phi_2}-e^{-2\phi_1},\qquad 
    \partial_{xt}\phi_2=e^{-\phi_1+\phi_2}-e^{-2\phi_1}. \label{eq1}
\end{gather}
It is clear that  $[E, D^{(-1)}_{vac}]=0$ and  henceforth  (\ref{eq1}) allows only zero vacuum solution, $\phi_i^{vac}=0, i=1,2$.
The next  model is  obtained for 
{$M=5, \;\; t= t_{-5}$} and eqn. of motion 
\begin{equation}
\begin{split}
        v_t=-e^{d^{-1}(2v-w)} &d^{-1}\left(e^{d^{-1}(-v+w)}d^{-1}\left[e^{-d^{-1}(v)}\left(c_{-3}-a_{-3}\right)\right]+e^{-d^{-1}(2v)}d^{-1}\left[e^{d^{-1}(w)}\left(c_{-3}-b_{-3}\right)\right]\right)\\
    &-2e^{-d^{-1}(2v)} d^{-1}\left(e^{d^{-1}(2v-w)}d^{-1}\left[e^{d^{-1}(w)}\left(c_{-3}-b_{-3}\right)\right]\right)
\end{split}
\end{equation}
and
\begin{equation}
\begin{split}
        w_t&=e^{d^{-1}(-v+w)} d^{-1}\left(e^{d^{-1}(2v-w)}d^{-1}\left[e^{-d^{-1}(v)}\left(c_{-3}-a_{-3}\right)\right]\right)\\&-2e^{-d^{-1}(2v)} d^{-1}\left(e^{d^{-1}(2v-w)}d^{-1}\left[e^{d^{-1}(w)}\left(c_{-3}-b_{-3}\right)\right]\right)
\end{split}
\end{equation}
where 
\begin{equation}
    \begin{split}
        a_{-3}&=-e^{d^{-1}w}d^{-1}\left[e^{-d^{-1}(2v)} d^{-1}\left(e^{d^{-1}(2v-w)}\right)\right]\\
        b_{-3}&=4e^{2d^{-1}(v-w)}d^{-1}\left[e^{-d^{-1}(v-w)}d^{-1}\left(e^{-d^{-1}(v-w)}\right)\right]\\
        c_{-3}&=e^{-d^{-1}v}d^{-1}\left[e^{-d^{-1}(v-w)} d^{-1}\left(e^{d^{-1}(2v-w)}\right)-2e^{d^{-1}(2v-w)} d^{-1}\left(e^{-d^{-1}(v-w)}\right)\right].\\
    \end{split}
\end{equation}
After a tedious but straightforward calculation  it can be   verified  that  $v=v_0$ e $w=w_0$ are indeed solutions of the equations of motion as expected  by extending the arguments  derived for  $A_2^{(2)}$   case in arguments  in (\ref{vac1}).

\section{ Comments  and Further Developments}

In this paper we  have studied  the construction of  integrable hierarchies  associated to the   $A_{2r}^{(2)}$  twisted affine  algebra 
We have discussed explicitly   the $A_2^{(2)}$ case as an example   and discussed in detail  the  algebraic construction of the positive   and negative sub-hierarchies.  
An important ingredient in classifying the  models  is the structure of the zero curvature in the vacuum configuration, ie., $A_{\mu}^{vac}$.  
These  algebraic quantities   contain  terms of different  $Q$ gradations.
The sub-hierarchies were shown to be    classified according to  i) a grading  structure  of algebraic nature ($Q$-gradation ) and ii) its  {\it zero} or {\it non zero} vacuum solution.  
We now argue that the  two  concepts can be put together  in terms of the  two-loop Kac-Moody algebra    proposed  many years ago in connection with conformal affine Toda models \cite{twoloop}, \cite{schwimmer}. Consider 
\br
\lb T^a_{m,r},  T^b_{n,s}\rb =if^{abc} T^c_{m+n, r+s}  +\kappa g_{ab}r\d_{r+s,0}\d_{m+n,0} +\tilde \kappa g_{ab}m\d_{m+n,0}\d_{r+s,0} \label{twoloop}
\er
where $\kappa$ and $\tilde \kappa $  denote  two central terms, $d$ and $\tilde d$,
 the   two derivation operators,
\br
\lb d, T^a_{m,r}\rb = m T^a_{m,r}, \qquad    \lb \tilde d, T^a_{m,r}\rb = r T^a_{m,r}.
\er
For the centerless version of (\ref{twoloop}) the  generators can be  realized  as 
\br T^a(z, w) = \sum_{m,r \in Z} T^a_{m,r} z^{-m} w^{-r} , \qquad d= z{{\pa }\over {\pa z}}, \qquad   \tilde d= w{{\pa }\over {\pa w}}. \label {discrete}
\er

Considering  $Q=6d+ (\mu_1+\mu_2)\cdot H$ of section 2 and defining $w=v_0$ to describe the second loop, under the new gradation   $\tilde Q = Q+  \tilde d  $ decomposes  the  affine algebra $\hat \lie = \oplus \tilde {\lie_a}$ and , 
\br
A_x^{vac} = E^{(1)} + v_0 (h_1^{(0)} + h_2^{(0)}) \in \tilde { \lie}_1.
\er
Moreover, all terms in $\Omega_{6m+1} \in \tilde {\lie}_{6m+1}$ in (\ref{O}) and $\G_{6m+5} \in \tilde {\lie}_{6m+5}$ in (\ref{G}) have the same grade according to the $\tilde Q$ gradation and so does the terms in 
$ A_{t_{6n\pm1}}^{vac} $ in (\ref{a-vac}) and $ A_{t_{-M}}^{vac} $ in (\ref{34}). 
The central extension $\kappa$  plays  a role to ensure  highest weight states $|i>$   and is responsible for introducing the field $\nu$  in (\ref{dress}). 
The role of the  second central term $\tilde \kappa$  is still not clear in such context and is a subject of interest in future developments.

As  a generalization of ref. \cite{assis} we have constructed  soliton solutions for  all models within the  $A_2^{(2)}$ hierarchy. 
Those  with  non zero vacuum  involves the construction of deformed vertex operators in terms of   the vacuum  parameter $v_0$.
A few simple examples  within the $A_2^{(1)}$ positive sub hierarchy (mKdV)  \cite{guilherme} shows that these  deformed solutions  satisfy   
Bäcklund transformation.  It would be interesting  to extend and understand the  structure of  (twisted) deformed   integrable defects in the lines of \cite{robertson}  and to consider algebras other then $A_{2r}^{(2)}$.

{\bf Acknowlegements}
JFG and AHZ thank CNPq and Fapesp for support. YFA thanks S\~ao Paulo Research Foundation (FAPESP) for financial support under grant $\# 2021/00623-4$ and GVL is supported by Capes.

\section{Appendix A}
Here we display the  solution for  $t_7$ time evolution   for the $A_2^{(2)}$ model.
\begin{equation}
    \begin{split}
        D^{(7)}&=E_{\a_1}^{(1)}+E_{\a_2}^{(1)}+E_{-(\a_1+\a_2)}^{(\frac{3}{2})}\\
        D^{(6)}&=v(h_1^{(1)}+h_2^{(1)})\\
         D^{(5)}&=-\frac{1}{3}(v^2+v_x)(E_{-\a_1}^{(1)}+E_{-\a_2}^{(1)})-\frac{1}{3}(v^2-2v_x)E_{(\a_1+\a_2)}^{(\frac{1}{2})}\\
        D^{(4)}&=-\frac{1}{3} \left(v^3-v_xv -v_{2x}\right)(E_{\a_1}^{(\frac{1}{2})}-E_{\a_2}^{(\frac{1}{2})})\\
          D^{(3)}&=-\frac{1}{9}(v^4+2 v_x v^2-2v_{2x} v-v_x{}^2-v_{3x})(h_1^{\frac{1}{2}}-h_2^{\frac{1}{2}})\\
          D^{(2)}&=\frac{1}{9}(4 v_{1x} v^3+2 v_{2x} v^2+4 v_{1x}^2v-2v v_{3x}-4 v_{1x} v_{2x}-v_{4x})(E_{-\a_1}^{(\frac{1}{2})}-E_{-\a_2}^{(\frac{1}{2})})\\
 D^{(1)}&=\frac{1}{81} (4 v^6-24v_x v^4-18v_{2x} v^3 + 12v^2v_x{}^2 +18v^2v_{3x}+72vv_xv_{2x}+3vv_{4x} v+4v_x{}^3-24v_xv_{3x}-9v_{2x}{}^2-3v_{5x})\\ &(E_{\a_1}^{(0)}+E_{\a_2}^{(0)})+
     \frac{1}{81} (4 v^6+12 v_x v^4-60 v_x{}^2 v^2 -36v^3v_{2x}-72v v_{2x} v_x12 v_{4x} v-32 v_x{}^3+27 v_{2x}{}^2\\&+30 v_{3x} v_x-18 v^2v_{3x}+6 v_{5x})E_{-(\a_1+\a_2)}^{(\frac{1}{2})}\\
  D^{(1)}&=\frac{1}{81} \left(4 v{}^6-24 v_x v{}^4-18 v_{2x} v{}^3+6v^2 \left(2 v_x{}^2+3 v_{3x}\right) +3v \left(24 v_x v_{2x}+v_{4x}\right)+4 v_x{}^3-24 v_x v_{3x}-3 \left(3 v_{2x}{}^2+v_{5x}\right)\right)\\ &E^{(1)}+\frac{1}{9} \left(4 v_x v{}^4-2 v_{2x} v{}^3-4 \left(2 v_x{}^2+v_{3x}\right) v{}^2+\left(v_{4x}-16 v_x v_{2x}\right) v-4 v_x{}^3+4 v_{2x}{}^2+6 v_x v_{3x}+v_{5x}\right)E_{-(\a_1+\a_2)}^{(\frac{1}{2})}\\
 D^{(0)}&= \frac{1}{81} (4v^7-42v_{2x}v^4-84 v_{1x}^2v^3+21(2v_{x} v_{2x}+ v_{4x})v^2+7(4 v_{x}^3+12v_{3x} v_{x} +9v_{2x}^2)v_1\\&+42v_{2x}(2 v_{1x}^2- v_{3x})-21v_{1x}v_{4x}-3v_{6x})(h_1^{0}+h_2^{0})
    \end{split}
\end{equation}

\begin{equation}
\begin{split}
      A_{t_7}^{vac}&=E_{\a_1}^{(1)}+E_{\a_2}^{(1)}+E_{-(\a_1+\a_2)}^{(\frac{3}{2})}+v_0(h_1^{(1)}-h_2^{(1)})-\frac{v_0^2}{9}[3(E_{-\a_1}^{(1)}+E_{-\a_2}^{(1)}+E_{(\a_1+\a_2)}^{(\frac{1}{2})})+3v_0(E_{-\a_1}^{(\frac{1}{2})}-E_{-\a_2}^{(\frac{1}{2})})\\
      &v_0^2(h_1^{\frac{1}{2}}-h_2^{\frac{1}{2}})]+\frac{4 v_0^6}{81}[E_{\a_1}^{(0)}+E_{\a_2}^{(0)}+E_{-(\a_1+\a_2)}^{(\frac{1}{2})}+v_0(h_1^{0}+h_2^{0})]
\end{split}
\end{equation}

\section{Appendix B}

 
For the $A_4^{(2)}$ case we find

\begin{gather}
    \mathbf{g_{10m}}=\{h_1^{(m)}+h_4^{(m)}, h_2^{(m)}+h_3^{(m)}\}\\
     \mathbf{g_{10m+1}}=\{E_{\a_1}^{(m)}+E_{\a_4}^{(m)},E_{\a_2}^{(m)}+E_{\a_3}^{(m)}, E_{-(\a_1+\a_2+\a_3+\a_4})^{(m+\frac{1}{2})}\}\\
      \mathbf{g_{10m+2}}=\{E_{(\a_1+\a_2)}^{(m)}-E_{(\a_3+\a_4)}^{(m)}, E_{-(\a_1+\a_2+\a_3)}^{(m+\frac{1}{2})}-E_{-(\a_2+\a_3+\a_4)}^{(m+\frac{1}{2})}\}\\
       \mathbf{g_{10m+3}}=\{ E_{(\a_1+\a_2+\a_3)}^{(m)}+E_{(\a_2+\a_3+\a_4)}^{(m)},E_{-(\a_2+\a_3)}^{(m+\frac{1}{2})}, 
        E_{-(\a_1+\a_2)}^{(m+\frac{1}{2})}+E_{-(\a_3+\a_4)}^{(m+\frac{1}{2})}\}\\
        \mathbf{g_{10m+4}}=\{E_{-\a_1}^{(m+\frac{1}{2})}-E_{-\a_4}^{(m+\frac{1}{2})}, E_{-\a_2}^{(m+\frac{1}{2})}-E_{-\a_3}^{(m+\frac{1}{2})}\}\\
        \mathbf{g_{10m+5}}=\{h_1^{(m+\frac{1}{2})}-h_4^{(m+\frac{1}{2})}, h_2^{(m+\frac{1}{2})}-h_3^{(m+\frac{1}{2})}\}\\
        \mathbf{g_{10m+6}}=\{E_{\a_1}^{(m+\frac{1}{2})}-E_{\a_4}^{(m+\frac{1}{2})}, E_{\a_2}^{(m+\frac{1}{2})}-E_{\a_3}^{(m+\frac{1}{2})}\}\\
        \mathbf{g_{10m+7}}=\{E_{-(\a_1+\a_2+\a_3)}^{(m+1)}+E_{-(\a_2+\a_3+\a_4)}^{(m+1)}, E_{(\a_2+\a_3)}^{(m+\frac{1}{2})}, 
        E_{(\a_1+\a_2)}^{(m+\frac{1}{2})}+E_{(\a_3+\a_4)}^{(m+\frac{1}{2})}\}\\
        \mathbf{g_{10m+8}}=\{E_{-(\a_1+\a_2)}^{(m+1)}-E_{-(\a_3+\a_4)}^{(m+1)},E_{(\a_1+\a_2+\a_3)}^{(m+\frac{1}{2})}-E_{(\a_2+\a_3+\a_4)}^{(m+\frac{1}{2})}\}\\
        \mathbf{g_{10m+9}}=\{E_{-\a_1}^{(m+1)}+E_{-\a_4}^{(m+1)},E_{-\a_2}^{(m+1)}+E_{-\a_3}^{(m+1)}, E_{(\a_1+\a_2+\a_3+\a_4})^{(m+\frac{1}{2})}\}
\end{gather}

Let 
\br
 A_0 =  v(h_1+h_4) + w(h_2+h_3), \qquad E^{(1)}= E_{\a_1}^{(0)}+E_{\a_2}^{(0)}+E_{\a_3}^{(0)}+E_{\a_4}^{(0)}+E_{-(\a_1+\a_2+\a_3+\alpha_4)}^{(\frac{1}{2})}
\er
and the  Kernel  of $E$   is given by
\br
        E^{(10m+1)}&= &E_{\a_1}^{(m)}+E_{\a_2}^{(m)}+E_{\a_3}^{(m)}+E_{\a_4}^{(m)}+E_{-(\a_1+\a_2+\a_3+\alpha_4)}^{(m+\frac{1}{2})} \label{k1} \\
        E^{(10m+3)}&=& E_{(\a_1+\a_2+\a_3)}^{(m)}+E_{(\a_2+\a_3+\a_4)}^{(m)}+E_{-(\a_2+\a_3)}^{(m+\frac{1}{2})}+ 
        E_{-(\a_1+\a_2)}^{(m+\frac{1}{2})}+E_{-(\a_3+\a_4)}^{(m+\frac{1}{2})} \label{k2} \\
        E^{(10m+7)}&= &E_{-(\a_1+\a_2+\a_3)}^{(m+1)}+E_{-(\a_2+\a_3+\a_4)}^{(m+1)}+ E_{(\a_2+\a_3)}^{(m+\frac{1}{2})}+
        E_{(\a_1+\a_2)}^{(m+\frac{1}{2})}+E_{(\a_3+\a_4)}^{(m+\frac{1}{2})} \label{k3} \\
         E^{(10m+9)}&=& E_{-\a_1}^{(m+1)}+E_{-\a_2}^{(m+1)}+E_{\a_3}^{(m+1)}+E_{-\a_4}^{(m+1)}+E_{(\a_1+\a_2+\a_3+\alpha_4)}^{(m+\frac{1}{2})} \label{k4}
    \er
$m \in Z$.  Following the  general pattern developed  for the $A_2^{(2)}$ case, we  expect to  obtain  positive sub-hierarchies  associated to each element of the Kernel  in (\ref{k1})-(\ref{k4})

\end{document}